# Influence of Photoemission Geometry on Timing and Efficiency in 4D Ultrafast Electron Microscopy

Simon A. Willis and David J. Flannigan[*]

*Department of Chemical Engineering and Materials Science, University of Minnesota, 421 Washington Avenue SE, Minneapolis, MN 55455, USA*

**Abstract:** Broader adoption of 4D ultrafast electron microscopy (UEM) for the study of chemical, materials, and quantum systems is being driven by development of new instruments as well as continuous improvement and characterization of existing technologies. Perhaps owing to the still-high barrier to entry, the full range of capabilities of laser-driven 4D UEM instruments has yet to be established, particularly when operated at extremely low beam currents (~fA). Accordingly, with an eye on beam stability, we have conducted particle tracing simulations of unconventional off-axis photoemission geometries in a UEM equipped with a thermionic-emission gun. Specifically, we have explored the impact of experimentally adjustable parameters on the time-of-flight (TOF), the collection efficiency (CE), and the temporal width of ultrashort photoelectron packets. The adjustable parameters include the Wehnelt aperture diameter ($D_W$), the cathode set-back position ($Z_{tip}$), and the position of the femtosecond laser on the Wehnelt aperture surface relative to the optic axis ($R_{photo}$). Notable findings include significant sensitivity of TOF to $D_W$ and $Z_{tip}$, as well as non-intuitive responses of CE and temporal width to varying $R_{photo}$. As a means to improve accessibility, practical implications and recommendations are emphasized wherever possible.

*Corresponding author
Email: flan0076@umn.edu
Office: +1 612-625-3867

## I. Introduction

Significant growth is presently occurring in the application of ultrafast electron microscopes (UEMs) employing femtosecond (fs) laser-driven cathodes. In addition to continued expansion of the studies of nanoscale materials phenomena,[1–7] fs laser-driven UEMs have been shown to be extremely well-suited for probing fundamental quantum behaviors of electron-photon interactions via photon-induced near-field electron microscopy (PINEM).[8–12] These and other UEM advances have benefited from electron-source and gun development; all transmission electron microscopy (TEM) gun types have now been successfully extended to laser-driven UEMs.[13–16] Indeed, the high degree of control over electron-beam trajectories and properties in fs laser-driven UEMs continues to open new research avenues, including the development of attosecond electron microscopy,[17–21] the probing of charge-carrier and plasma dynamics (*i.e.*, charge dynamics electron microscopy or CDEM),[22,23] and the exploration of pulsed-beam radiation damage.[24–26]

While key pulsed electron-beam parameters, such as brightness and coherence, have rightfully received substantial attention, far less is presently known about the stability of fs laser-driven photoelectron beams in UEMs. Gaining an understanding of the factors that influence stability has significant implications with respect to achievable resolutions, sensitivities, and quality of data and information that can be gathered. Despite the present dearth of data, factors influencing conventional electron-beam stabilities are hypothesized to also impact the photoemitted beams (*e.g.*, tip contamination or structural modification).[16,27,28] This presents an opportunity to explore electron-source materials and configurations that provide improved stability without significantly compromising other application-driven properties.

Unconventional photoemitting geometries have shown promise for increasing versatility and stability, without significantly compromising usability.[29] For example, we showed that off-axis photoemission from the Wehnelt-aperture surface of a thermionic electron gun can readily produce pulsed beams with ~10 pA currents that are stable, usable, and robust.[27] However, the altered geometries cause associated shifts in electron-packet time-of-flight (TOF) that manifest in the onset of dynamics. Shifts on the order of tens of picoseconds have been observed, comparable to shank photoemission from tapered photocathodes.[30,31]

Thus, to gain a better understanding of the electron trajectories of off-axis configurations and, more importantly, the impact on critical parameters, such as temporal distribution and collection efficiency (CE), we have performed particle-tracing simulations for a UEM gun equipped with a Wehnelt electrode. We systematically varied key component dimensions and geometries, including the Wehnelt aperture diameter, the off-axis position of the UV laser on the aperture, and the set-back position of the cathode relative to the aperture plane. The results show that, in addition to replicating experimentally observed shifts in TOF, the off-axis configuration creates an energy-filtering effect, where a narrower range of the overall electron distribution passes through the limiting X-ray aperture before entering the illumination system. While this reduces CE relative to on-axis geometries, and thus negatively impacts beam current, it also reduces the energy spread (and, thus, the temporal spread). We conclude by providing our perspective on the implications of these findings.

## II. Methods

*1. Gun architecture and photoemission properties.* The elements and dimensions of the Tecnai Femto UEM gun region, as well as the software tools and simulation methods used here, have also been used to study on-axis photoelectron-beam properties.[32,33] Pertinent details are again provided here for convenience. The exact architecture and dimensions for the gun region of the Thermo Fisher/FEI Tecnai Femto UEM comprised the physical elements of the simulations (Fig. 1).[a] For reference, the base instrument is a Tecnai $G^2$ T20 TEM. The key parameters of interest were the Wehnelt aperture diameter ($D_W$, varied), the aperture-to-cathode-tip distance ($Z_{tip}$, varied), and the centroid of the photoemitting spot (experimentally dictated by the Gaussian laser-beam spot) relative to the Wehnelt aperture edge ($R_{photo}$, varied). Here, a Gaussian photoemitting spot of 50 µm (fwhm) and a Gaussian temporal pulse duration of 300 fs (fwhm) were used throughout. The photocathode tip diameter was fixed at 0.2 mm in order to match prior experimental work.[31] In UEM mode, the Wehnelt triode in the Tecnai Femto acts as a simple, fixed electrostatic lens.[34,35]

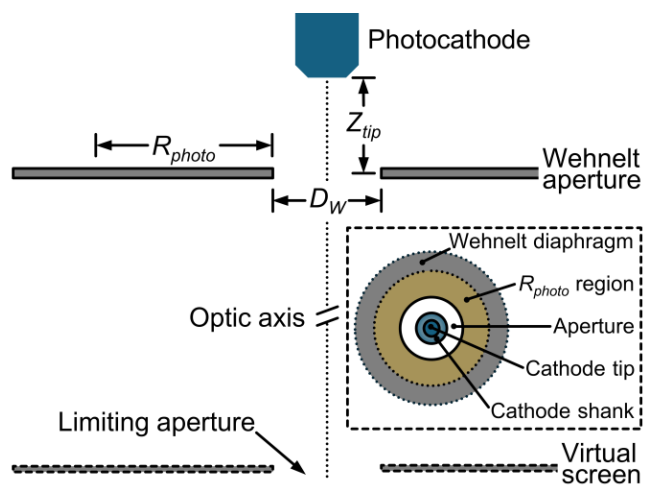

**Figure 1.** Schematic of the simulation elements, geometries, and dimensions. Pertinent elements are labeled. $Z_{tip}$ refers to the distance from the tapered, truncated cathode tip surface to a parallel surface that bisects the aperture plane. $R_{photo}$ refers to the distance of the centroid of the Gaussian photoemitting spot to the aperture edge – a representative range covered by specific $R_{photo}$ centroid

---

[a] Dimensions and geometries provided by Erik Kieft of Thermo Fisher Scientific.

positions is indicated. $D_W$ refers to the Wehnelt-diaphragm aperture diameter. The virtual screen marks the termination plane of the simulated trajectories and is the endpoint for determining electron time-of-flight (TOF). The limiting aperture (equivalent to the fixed X-ray aperture) determines the photoelectron collection efficiency (CE). (Inset) Schematic of the underside of the Wehnelt triode as viewed along the optic axis, with pertinent elements and regions labeled. Dotted circles denote virtual boundaries, while solid circles are physical boundaries.

Particle tracing simulations were conducted using General Particle Tracer (GPT, Pulsar Physics). Cylindrically symmetric field maps for the Tecnai Femto gun architecture were calculated with Poisson Superfish.[36,37] GPT is used to solve the relativistic equations of motion with a fifth-order embedded Runge-Kutta solver and to calculate the Lorentz force acting on the particle. Poisson Superfish consists of a finite element method used to solve Poisson's equation for electrostatics. For all simulated configurations, trajectories of $5 \times 10^4$ non-interacting photoelectrons emitted from a Gaussian-shaped spot were tracked, whether from the photocathode or from the Wehnelt aperture. Initial kinetic energy spreads were calculated using Fermi-Dirac statistics and an energy step function based on a work-function value ($\Phi$) of 4.2 eV and an incident photon energy ($h\nu$) of 4.8 eV; note that previous work investigated the effects of varying ($h\nu - \Phi$) in the Tecnai Femto,[32] and so a fixed value for $\Phi$ was used here in order to focus on the effects of the off-axis geometry. Initial momenta were assigned according to an azimuthally integrated $\cos\theta$ distribution.[32,33]

*2. Representative simulation results.* The key output of the simulations is the arrival time of each photoelectron at the limiting aperture in the virtual screen positioned 0.35 m along the optic axis from the photocathode tip surface (see Fig. 1). The time elapsed between launch from the photoemitting surface and arrival at the limiting aperture is defined as the time-of-flight (TOF). The number of photoelectrons passing through the limiting aperture vs. the total number launched ($5 \times 10^4$) is defined as the collection efficiency (CE). Figure 2 displays representative histograms of TOF of photoelectrons launched from the on-axis photocathode and from the off-axis Wehnelt aperture surface. All comparisons of parameters of interest between the two photoemitting surfaces are made from such data.

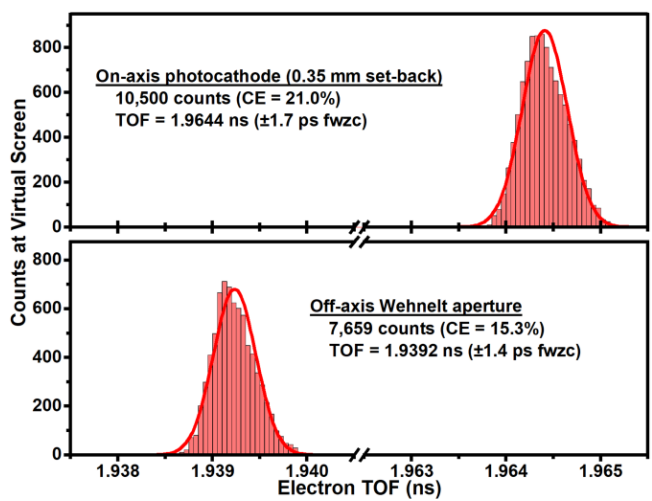

**Figure 2.** Histograms of simulated electron counts reaching the limiting aperture in the virtual screen vs. electron TOF for the on-axis 0.2 mm photocathode (top plot) and the off-axis Wehnelt aperture (bottom plot). The associated CEs and TOFs are noted (fwzc = full-width at zero-counts). The bin size is 50 fs, and the histograms are fit with normal distributions. Here, $Z_{tip}$ = 0.35 mm, $D_W$ = 1.0 mm, and $R_{photo}$ = 0.125 mm (*i.e.*, the centroid is 0.625 mm from the optic axis – see Fig. 1).

For the representative case in Figure 2, the CEs and TOFs were 21.0% and 1.9644 ns for the on-axis geometry vs. 15.3% and 1.9392 ns for the off-axis geometry (see caption for details).

Thus, the peak positions and full-widths at zero-counts (fwzc) of the normal distribution functions fit to the histograms indicate the TOF for electrons emitted from the Wehnelt aperture arrive 25.2 ± 3.1 ps before those from the photocathode (error is propagated from the fwzc values). Note that this agrees reasonably well with experiments.[31] As will be shown below, simple relative positions of the photoemitting surfaces only partially account for the TOF shifts. Further, the fwzc is narrower for photoemission from the off-axis aperture vs. the on-axis cathode. Generally, this is due to effective selection of electrons spanning a narrower range of transverse momenta by the limiting aperture due to the off-axis geometry.[31] This effect also manifests in the packet temporal widths (see Fig. 5).

Note that the histogram distributions show a slight but consistent asymmetry, with a bunching at earlier electron TOF values and a spreading at later values. The origins of this chirped distribution can be traced back to the initial kinetic-energy and momenta distributions; chirped electron packets in UEM have been experimentally observed with PINEM and are a common trait arising from Coulombic interactions.[38–41] However, because the effect is small relative to the fwzc values, we chose to focus mainly on the differences in TOF and CE of the two geometries. We therefore have approximated the histograms to normal distributions throughout. Overall, the behaviors shown in Figure 2 are qualitatively consistent across all simulations – photoelectrons from the Wehnelt aperture arrive earlier and with a narrow temporal distribution than those from the on-axis cathode.[31] As will be shown, however, there are configurations under which the behaviors deviate from the general trends.

### III.  Results and Discussion

Figure 3 summarizes the results of simulations comparing the TOFs of photoelectrons emanating from the Wehnelt aperture surface and the cathode surface as a function of $Z_{tip}$ for three values of $D_W$ (0.3, 0.7, and 1.0 mm). Note that the 0.7-mm size was chosen because it is commonly used in TEMs with thermionic electron guns, while the 1.0-mm size was chosen because our lab used Wehnelt apertures of this diameter in order to obtain higher UEM beam currents when on-axis photoemission was primarily used. The 0.3-mm size was chosen in order to study behaviors for apertures that would enable photoemission reasonably close to the optic axis. However, based on our experience, it is unlikely that reliably stable thermionic beams could be generated with an aperture of this size, thus negating use of thermionic and photoemission modes in a single instrument.[27]

The results in Figure 3 are reported as the relative TOF (*r*-TOF) of Wehnelt-aperture photoelectrons. Here, *r*-TOF is defined as the average cathode-photoelectron TOF subtracted from the average Wehnelt-aperture TOF (*i.e.*, $\overline{TOF}_{Wehnelt} - \overline{TOF}_{cathode}$). For example, an *r*-TOF of -25 ps in the figure indicates the associated geometry causes Wehnelt-aperture photoelectrons to arrive at the virtual screen 25 ps *before* those emitted from the cathode, as is the case for the representative simulation results in Figure 2. Also, negative $Z_{tip}$ values denote a configuration where the photocathode is protruding from the Wehnelt cap (*i.e.*, the cathode tip surface extends *below* the aperture mid-plane). Note that positive $Z_{tip}$ values are the norm experimentally (*e.g.*, $Z_{tip} \cong 0.35$ mm is commonly used in the UEM lab at Minnesota).[34]

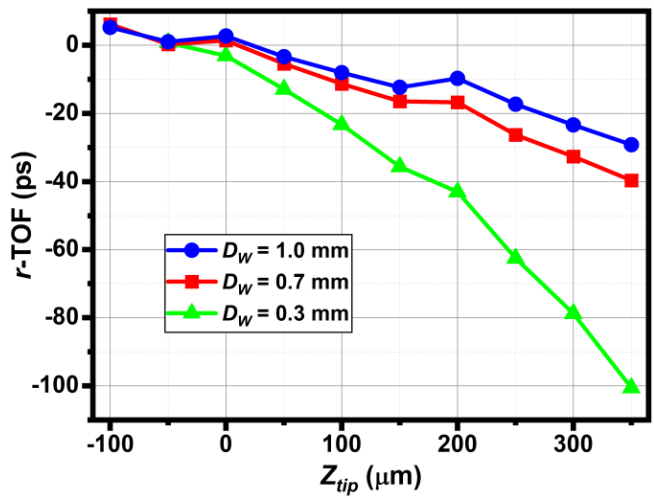

**Figure 3.** Relative TOF ($r$-TOF; $\overline{TOF}_{Wehnelt} - \overline{TOF}_{cathode}$) for Wehnelt-aperture photoelectrons vs. $Z_{tip}$ for three values of $D_W$ (blue circles = 1.0 mm; red squares = 0.7 mm; green triangles = 0.3 mm). Note that negative $Z_{tip}$ values denote a condition where the cathode tip surface is protruding outside the Wehnelt cap (*i.e.*, below the Wehnelt aperture mid-plane). The lines between data points are to guide the eye.

Before discussing the noteworthy trends in Figure 3, it is important to point out that the simulation results showed that $r$-TOF for all tested combinations of $D_W$ and $Z_{tip}$ is *almost entirely independent* of $R_{photo}$. That is, changing $R_{photo}$ has virtually no effect on $r$-TOF. For example, for $D_W = 0.7$ mm and $Z_{tip} = 0.35$ mm, $r$-TOF is an almost constant -35 ps for all $R_{photo}$ values ranging from 50 to 185 µm (*i.e.*, from 0.4 to 0.535 mm from the optic axis). This is not altogether surprising owing to the large difference between the usable range of $R_{photo}$ values – the maximum of which is ~0.2 mm – and the distance to the virtual screen (350 mm). (Note that CE drops to zero beyond $R_{photo} \cong 0.2$ mm for all simulated configurations.) Accordingly, a constant $R_{photo}$ value applies to all data shown in Figure 3.

As for the trends, first, regardless of $D_W$, $r$-TOF generally becomes more negative as the cathode set-back distance ($Z_{tip}$) increases (*i.e.*, photoelectrons from the Wehnelt aperture generally arrive earlier). Second, the trend to larger negative values of $r$-TOF is more pronounced for small Wehnelt apertures, especially at larger values of $Z_{tip}$. For example, $r$-TOF for $D_W$ = 1.0, 0.7, and 0.3 mm is -12.4, -16.5, and -35.6 ps, respectively, at $Z_{tip}$ = 150 µm, while it is -29.2, -39.7, and -100.5 ps, respectively, at $Z_{tip}$ = 350 µm. Thus, the difference between the maximum and minimum $r$-TOF values increases from 23.2 ps at $Z_{tip}$ = 150 µm to 71.3 ps at $Z_{tip}$ = 350 µm. This indicates that, while simple relative positions of the emitting surfaces partially account for the overall trend to earlier arrival time, the aperture size itself has a more dramatic effect. This is especially apparent owing to the lack of dependence on $R_{photo}$.

The trends in Figure 3 generally arise from the Wehnelt electrode acting as a fixed electrostatic lens in the Tecnai Femto when the cathode is not heated to the thermionic emission threshold (*i.e.*, while in UEM mode).[34] Despite the lack of independent bias control, electrons emitted from the photocathode still experience a lensing effect, with beam crossovers present for certain electron-gun configurations (*i.e.*, for certain combinations of $D_W$ and $Z_{tip}$).[32–34] This lensing effect increases the pathlength of such electrons, with the impact of the effect increasing with increasing transverse field strength in the plane of the Wehnelt aperture. Indeed, the impact of Wehnelt bias on electron TOFs has been demonstrated in a JEOL-based UEM.[30,38,42] For unbiased or constant-bias electrodes, the field strength increases nonlinearly with decreasing $D_W$, which is in agreement with the trends shown in Figure 3.[32–34] At negative $Z_{tip}$ values (*i.e.*, when the cathode tip surface lies below the aperture plane), trajectories of photoelectrons emitted from the cathode are impacted to a similar degree regardless of $D_W$. However, photoelectrons emitted from cathodes at relatively large $Z_{tip}$ positions and propagating near the optic axis will arrive at larger radial positions in the Wehnelt-aperture plane. Thus, such electrons experience a stronger field force and traverse paths with greater ellipticity compared to those with the same trajectories but emitted from cathodes at smaller $Z_{tip}$ values.

Unlike $r$-TOF, CE is quite sensitive to $R_{photo}$. This sensitivity is illustrated in Figure 4, which shows the CE response as a function of $R_{photo}$ for each aperture size at a select $Z_{tip}$ of 350 µm. Several noteworthy trends are apparent. First, the smallest aperture diameter produces the largest overall CE ($CE_{max} = 0.22$) – this is the case when making comparisons at common $Z_{tip}$ values, as well as when conditions are individually optimized for CE for each Wehnelt aperture diameter. Further, an overall trend toward lower $CE_{max}$ with increasing $D_W$ is observed, from 0.22 for $D_W = 0.3$ mm to 0.16 for $D_W = 1.0$ mm. Second, $CE_{max}$ is observed to move toward larger $R_{photo}$ values with increasing $D_W$, from 62.6 µm for $D_W = 0.3$ mm to 94.9 µm for $D_W = 1.0$ mm. Third, peak fitting reveals that the CE responses for each $D_W$ have nearly the same fwhm (~50 µm) to within ~4%. This third observation also shows that, regardless of $D_W$, the CE drops to 50% of $CE_{max}$ when the laser spot is translated approximately 25 µm to either side of the $CE_{max}$ position. This has implications with respect to photoelectron beam-current stability; in addition to potential spatial variations in quantum efficiency of the photoemitter, laser-beam pointing stability can be a non-trivial source of photobeam instability.

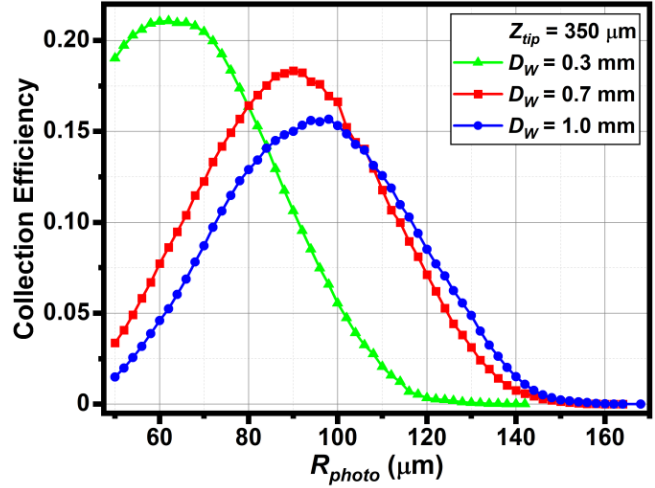

**Figure 4.** Simulated behavior of photoelectron collection efficiency (CE) when varying $R_{photo}$ for three different Wehnelt aperture sizes ($D_W$) at a fixed set-back position ($Z_{tip}$) of 350 µm. Green triangles are the 0.3-mm aperture data, red squares are the 0.7-mm aperture data, and the blue circles are the 1.0-mm aperture data. The lines between data points are to guide the eye.

It is important to note that, for the data shown in Figure 4, the inner aperture edge is set as $R_{photo} = 0$ for each $D_W$. The data begin at $R_{photo} = 50$ µm because the simulated Gaussian laser-spot size is 50 µm fwhm. Because each aperture is centered on the optic axis, which itself is defined as a line extending vertically from the center of the conventional photocathode down the microscope column (see Fig. 1), the inner aperture edge for each $D_W$ is at a different radial position relative to the optic axis. We refer to this parameter as $R_{photo\text{-}optic}$ to differentiate it from the normalized $R_{photo}$ parameter and to specify that it is with respect to the optic axis. Accordingly, $CE_{max}$ occurs at $R_{photo\text{-}optic} = 213, 440,$ and 595 µm for $D_W = 0.3, 0.7,$ and 1.0 mm, respectively, in Figure 4 (*i.e.*, for $Z_{tip} = 350$ µm). This is important to note because the field lines comprising the field maps are influenced by $D_W$ and thus are at least partly responsible for the observed trends. Note that full exploration of the entire parameter space studied here indicates that the $CE_{max}$ values in Figure 4 are the overall largest values in the system. Also note that we have confined our simulations to the experimentally accessible parameter space, as gleaned from experiments.[31]

In addition to the point about the sensitivity of *CE* to $R_{photo}$ and the practical implications for photobeam stability, Figure 4 also shows that $CE_{max}$ is rather insensitive to $D_W$. Based on the results, it appears that experimentally one can expect to achieve a similar optimum collection efficiency (~0.20) regardless of Wehnelt aperture size by optimizing the $R_{photo\text{-}optic}$ position for the given aperture. Further, the $R_{photo\text{-}optic}$ peak-CE position becomes significantly different across aperture diameters only when going to relatively small values of $D_W$. Thus, modest changes in laser alignment suffice when switching between apertures having practically useful diameters. Critically, however, the calculated field maps show that a strong field gradient exists near the aperture edges. This causes the photoelectrons emitted nearer to the aperture to be strongly deflected away from the optic axis and ultimately to not reach the virtual screen. This is reflected in the value of CE decreasing as the $R_{photo}$ position moves closer to the aperture edge.

Finally, in order to understand the impact of the off-axis photoemission geometry on UEM temporal resolution, we determined the photoelectron TOF distribution over a range of $R_{photo}$ positions, the results of which are summarized in Figure 5. We report the photoelectron packet duration as the full-width at tenth-max (fwtm) of a Gaussian distribution. As with *r*-TOF and CE, non-intuitive behaviors are observed for the temporal duration that can be rationalized by understanding how the specific fields in the gun region influence photoelectron trajectories. Overall, the behavior of the fwtm vs. $R_{photo}$ is mostly insensitive to $D_W$; we chose to discuss the $Z_{tip}$ = 350 µm results here, but the trends generally hold for other cathode set-back values. For each aperture, the fwtm is approximately 700 fs nearest the aperture edge (*i.e.*, nearest the optic axis) and gradually increases with increasing $R_{photo}$ to approximately 120 µm before showing relatively rapid declines in duration beyond this point. While the lowest reduced values at higher $R_{photo}$ are ~500 fs, the 0.7-mm aperture data does contain some points well below this (~100 fs).

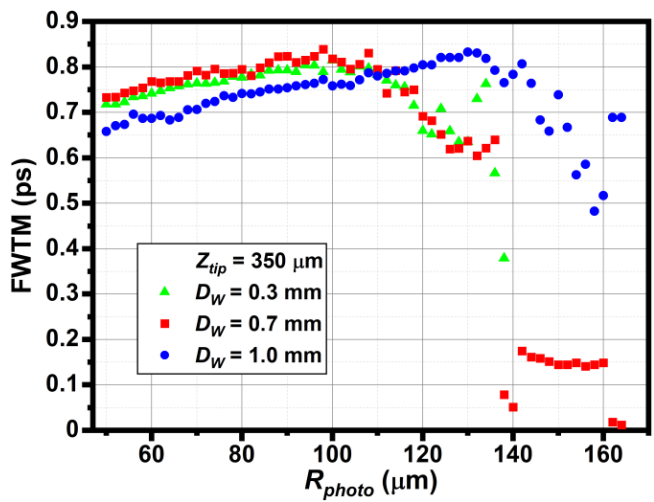

**Figure 5.** Simulated photoelectron packet temporal width (full-width at tenth-max, fwtm) as a function of laser-spot position ($R_{photo}$) on the Wehnelt aperture for the select $Z_{tip} = 350$ µm set-back position. Green triangles are the 0.3-mm aperture data, red squares are the 0.7-mm aperture data, and the blue circles are the 1.0-mm aperture data. Note that the packet temporal duration is the distribution of photoelectron TOFs that reach the virtual screen.

In order to understand the trends in Figure 5, one must consider both the behavior of CE with $R_{photo}$ (see Fig. 4) as well as details of the calculated field maps. First, recall that CE actually peaks at positions away from the aperture edge (*i.e.*, away from the optic axis). Indeed, the 0.7- and 1.0-mm apertures have peak efficiencies at $R_{photo}$ positions between 90 and 100 µm. Additionally, the CE behavior is mostly symmetric about the peak positions. (While the 0.3-mm aperture peak CE is significantly shifted toward the aperture edge, it also appears to have a mostly symmetric response.) The fall-off in fwtm values beyond $R_{photo}$ ~120 µm is due to an effective energy filtering effect of the photoelectron distribution. That is, the field lines are such that only a select few photoelectrons have trajectories that allow them to reach the virtual screen (as reflected by the relatively low CE in this region). These photoelectrons span a narrow range of kinetic energies relative to the initial distribution. Thus, the resultant fwtm is reduced. Conversely, while the CE is also relatively low close to the aperture edge, photoelectrons spanning a broader distribution of kinetic energies are admitted into the limiting x-ray aperture. Thus, the resultant fwtm is comparable to that at the peak CE positions. Practically, this illustrates a potential simple approach to improving UEM temporal resolution by taking advantage of the serendipitous energy-filtering effect, though at the expense of total beam current.

In conclusion, we have used particle tracing simulations and calculated field maps of the precise architecture of a UEM electron gun to study the behavior of photoelectron trajectories for unconventional off-axis emission geometries. The results reveal potentially useful trends that can act as a guide to experimental work in terms of optimization of electron-beam properties and hardware configuration in minimally modified instruments. Additionally, the results further clarify the operational subtleties and limits of thermionic-based UEMs, and we expect other such findings to emerge as work progresses. Ultimately, a complete mapping of the limits of such UEMs will be done, with combined simulation and experimental verification continuing to be critical to this effort.

**Author contributions**

S. A. W. contributions were data curation, formal analysis, investigation, methodology, software, validation, visualization, writing – original draft. D. J. F. contributions were conceptualization, data curation, funding acquisition, methodology, project administration, resources, supervision, visualization, writing – original draft, writing – review and editing.

**Conflicts of interest**

There are no conflicts to declare.

**Acknowledgements**

This material is based on work supported by the U.S. Department of Energy, Office of Science, Office of Basic Energy Sciences under Award No. DE-SC0023708. This work was supported partially by the National Science Foundation through the University of Minnesota MRSEC under Award Number DMR-2011401.